\documentclass[fleqn,10pt]{wlscirep}
\title{A stand-alone fiber-coupled single-photon source}

\author[1]{Alexander Schlehahn}
\author[1]{Sarah Fischbach}
\author[1]{Ronny Schmidt}
\author[1]{Arsenty Kaganskiy}
\author[1,2]{Andr\'e Strittmatter}
\author[1]{Sven Rodt}
\author[1,*]{Tobias Heindel}
\author[1]{Stephan Reitzenstein}
\affil[1]{Institut f\"ur Festk\"orperphysik, Technische Universit\"at Berlin, 10623 Berlin, Germany}
\affil[2]{Present address: Abteilung f\"ur Halbleiterepitaxie, Otto-von-Guericke Universit\"at, 39106 Magdeburg, Germany}

\affil[*]{tobias.heindel@tu-berlin.de}

\begin{abstract}
In this work, we present a stand-alone and fiber-coupled quantum-light source. The plug-and-play device is based on an optically driven quantum dot delivering single photons via an optical fiber. The quantum dot is deterministically integrated in a monolithic microlens which is precisely coupled to the core of an optical fiber via active optical alignment and epoxide adhesive bonding. The rigidly coupled fiber-emitter assembly is integrated in a compact Stirling cryocooler with a base temperature of 35\,K. We benchmark our practical quantum device via photon auto-correlation measurements revealing $g^{(2)}(0)=0.07 \pm 0.05$ under continuous-wave excitation and we demonstrate triggered non-classical light at a repetition rate of 80\,MHz. The long-term stability of our quantum light source is evaluated by endurance tests showing that the fiber-coupled quantum dot emission is stable within 4\% over several successive cool-down/warm-up cycles. Additionally, we demonstrate non-classical photon emission for a user-intervention-free 100-hour test run and stable single-photon count rates up to 11.7\,kHz with a standard deviation of 4\%. 
\end{abstract}
\begin{document}

\flushbottom
\maketitle

\thispagestyle{empty}

\section*{Introduction}

Solid-state based quantum-light sources are elementary building blocks for future photonic quantum networks \cite{Aharonovich2016}, quantum information processing \cite{Kok2007} and quantum metrology \cite{Giovannetti2004}. To date, the performance of non-classical light sources has reached such a high level, that the development of user-friendly devices for applications can be pursued. In particular, single-photon sources (SPSs) based on semiconductor quantum dots (QDs) show close to ideal properties in terms of the quantum nature of emission \cite{Aharonovich2016}, and sophisticated excitation schemes including strict resonant excitation enabled proof-of-principle demonstrations of photonic cluster-state generation \cite{Schwartz2016a} and boson sampling \cite{Wang2017,Loredo2017} in laboratory environments. Furthermore, field experiments using QD-based SPSs have been employed for proof-of-principle quantum key distribution experiments \cite{Rau2014}, but still suffered from complex and bulky closed-cycle pulse-tube coolers and complex light extraction via free-space optics. Against this background, it is clear that fiber-coupling and robust packaging of practical quantum-light emitting devices is highly desirable for taking steps beyond proof-of-principle experiments. First steps in this direction utilized fiber-coupled QD-SPSs requiring fiber-bundles containing about 600 individual fiber cores to spatially post-select single emitter in a non-deterministic sample layout \cite{Xu2007}. Non-deterministic device approaches were also employed for the fiber-coupling of the single-photon emission from QDs embedded in as-grown planar samples \cite{Kumano2016} or microcavity structures based on photonic crystals \cite{Lee2015,Daveau2017} and micropillars with oxide aperture \cite{Haupt2010,Snijders2017}. Moreover, the direct fiber-coupling of nitrogen-vacancies in nano-diamonds \cite{Schroeder2011a} and single QDs embedded in nanowires \cite{Cadeddu2016} was realized using less practical pick-and-place techniques based on micromanipulators in a scanning electron microscope (SEM). All of these previous reports used bulky cryostats to operate QDs at cryogenic environments, which hindered the development of user-friendly devices for real-world applications.

In this work, we report on a stand-alone and user-friendly quantum light source based on a fiber-coupled QD integrated within a compact Stirling cryocooler. The QD is deterministically embedded within a monolithic microlens via in-situ electron-beam lithography and precisely coupled to an optical fiber by employing an optical alignment-process and epoxide adhesive bonding at room-temperature. This concept allows for a high degree of control and reproducibility for the fabrication of stand-alone SPSs with pre-defined emission properties - features which are a key requirements for the upscaling of fiber-coupled quantum networks. To benchmark our source, we conduct photon auto-correlation measurements under continuous wave and pulsed optical excitation. Additionally, we demonstrate a high stability of the fiber-coupled single-photon emission over several cool-down/warm-up cycles as well as user-intervention-free long-term test runs, demonstrating the capability of our single-photon unit for autonomous operation in future photonic quantum networks.

\section*{Results}

\subsection*{Fiber-coupling of deterministic QD microlenses}

\begin{figure}[t]
\centering
\includegraphics[width=\linewidth]{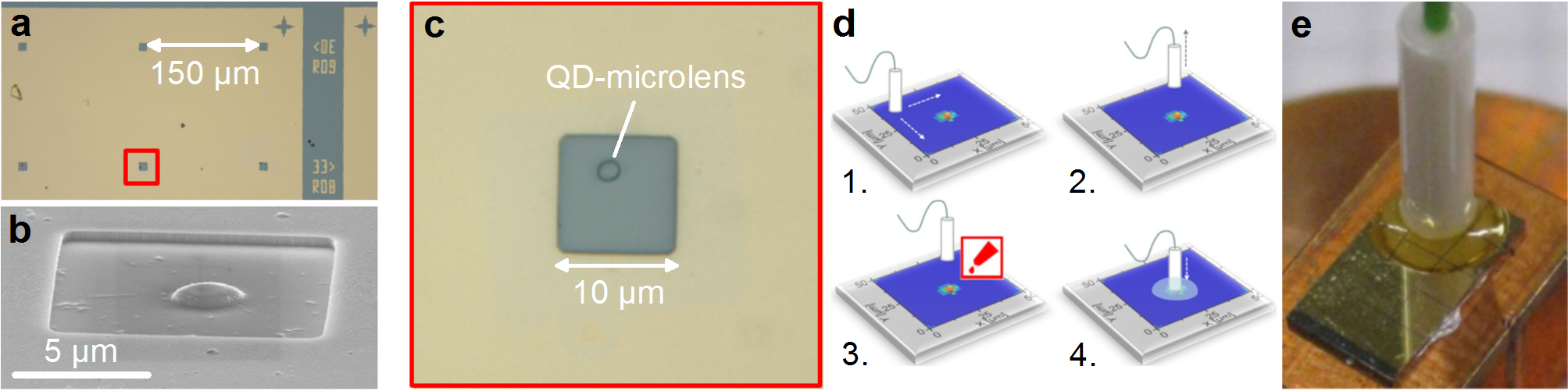}
\caption{Fabrication of fiber-coupled QD microlenses. (a) Microscope image of the sample surface before fiber-coupling. A gold-mask contains arrays of apertures with a pitch of 150\,$\mu$m. Each target aperture contains a single deterministically fabricated QD-microlens. Marker structures allow for unambiguous identification of target apertures and microlenses. (b) SEM image of a deterministic single-QD microlens fabricated via 3D in-situ electron-beam lithography and reactive ion etching. (c) Microscope image of a single aperture (dimensions: $10\mu\rm{m}\times10\mu\rm{m}$) containing a microlens deterministically fabricated above a pre-selected QD. Suitable QD-microlenses are pre-characterized using standard micro-photoluminescence spectroscopy at 10\,K. 
(d) Illustration of the room-temperature fiber-coupling process: 1. Fiber-scan across sample surface and monitoring of GaAs-bandgap emission within the gold apertures excited by 651\,nm laser. Emission of the bandgap is only visible above apertures and markers. 2. Precise alignment of the fiber above a precharacterized target aperture and lifting of the fiber by about 5\,mm. 3. Attaching a small drop of epoxide adhesive to the fiber-ferrule. 4. Lowering of the fiber to its previous position and monitoring of GaAs emission during hardening ($\approx2\,$hours). (e) Photograph of a fiber-coupled QD sample after the process illustrated in (d) showing the fiber ferrule glued to the sample.}
\label{Fig_1}
\end{figure}

The stand-alone SPS is based on a wafer grown by metal-organic chemical vapour deposition on GaAs-(001) substrate. After a 300\,nm thick GaAs buffer layer, a distributed Bragg reflector consisting of 23 pairs of $\lambda/4$-thick Al$_{0.9}$Ga$_{0.1}$As/GaAs is deposited. Next, 65\,nm of GaAs, a single layer of self-organized InAs QDs and a 420\,nm thick GaAs capping layer are grown. For the optical alignment procedure of the fiber core to a QD-microstructure, a 80\,nm-thick gold mask containing $10\,\mu\rm{m}\times10\,\mu$m apertures with a pitch of 150\,$\mu$m and corresponding markers (for unambiguous identification) is defined via optical (UV) lithography (cf. Figure \ref{Fig_1}\,a). Subsequently, the sample is spin-coated with a 100\,nm-thick layer of the electron-beam resist AR-P 6200 (CSAR 62)\cite{Kaganskiy2016} and transferred to a low-temperature cathodoluminescence lithography (CLL) system. The latter consists of a SEM with integrated liquid-helium flow-cryostat and extensions for optical spectroscopy as well as electron-beam lithography \cite{Gschrey2015}. Using the 3D in-situ electron beam lithography at 10\,K \cite{Gschrey2013,Gschrey2015b}, we define single deteministic QD-microlenses in the gold apertures. Afterwards the sample is transferred out of the CLL system and the resist is developed. This leaves the cross-linked resist above target QDs acting as lens-shaped etch masks for the subsequent etch step using reactive-ion-enhanced plasma etching (etch depth: 420\,nm). This directly transfers the lens-profile into the semiconductor material (i.e. GaAs). As a result of the described fabrication process, our sample is covered by a gold mask containing apertures with one deterministic QD-microlens (diameter: 2.4\,$\mu$m) each (cf. Figure \ref{Fig_1}\,b and c), while all other (non-selected) QDs inside apertures were removed by the dry-etching process. Using micro-photoluminescence spectroscopy at 10\,K, suitable QD-microlenses are selected according to their PL intensity. 

For the precise coupling of a photonic QD-microstructure to an optical fiber, we developed an alignment technique which combines deterministic in-situ QD-device fabrication with a robust optical coupling of the fiber core using epoxide adhesive bonding at ambient conditions (room temperature, no vacuum). In the developed process we scan a multi-mode fiber with 50\,$\mu$m core-diameter embedded in a ceramic ferrule in short distance (see below) across the sample surface using a 3D closed-loop piezo-stage, as illustrated in Figure \ref{Fig_1}\,d. The light of a CW laser ($\lambda=\rm{671}\,$nm) coupled into the fiber is reflected at the gold mask. If the fiber core is located above an aperture, charge-carriers are generated inside the semiconductor material and the associated photoluminescence of the GaAs bandgap (around 870\,nm at 300\,K) can easily be detected using a spectrometer attached to the output of the scanning fiber. The pitch of 150\,$\mu$m between apertures in combination with alignment markers allows us to precisely locate target apertures containing a single QD-microlens, which was pre-characterized via $\mu$PL. The height of the scanning fiber is thereby adjusted by observing the luminescence of the GaAs-bandgap in the following way: If the fiber ferrule is gently pressed to the sample surface, the resulting strain leads to a slight spectral shift of the bandgap emission. Moving the fiber back to the position of the unstrained GaAs-bandgap defines the point of physical contact.

After locating the target aperture, the fiber is lifted by about 5\,mm normal to the sample surface. At this point, a small drop of epoxide adhesive is attached to the fiber-ferrule and the fiber is lowered to its previous position. The final position has to be reached within the epoxide pot life of about 5\,minutes. Until the hardening process is completed ($\approx2\,$hours), the emission of the GaAs bandgap is monitored continuously to detect a possible misalignment between fiber-core and aperture. To test the accuracy of our fiber-coupling procedure, we repeated the process described above for fiber-core diameters of 25\,$\mu$m and 9\,$\mu$m. In all cases we were able to precisely locate the positions of the apertures, suggesting an alignment accuracy better than $9\,\mu$m. 

The presented approach for fiber-coupling has several advantages: Firstly, the complete alignment and curing process is performed under ambient conditions, and neither requires cryogenic temperatures for QD identification nor heating or UV illumination during hardening of the adhesive. This makes our process robust, easy to use and economic, and still enables a high accuracy of the alignment.
Secondly, the fiber is embedded in a ceramic ferrule which increases the bond area and hinders the fiber from being torn off by torsional forces appearing when handling the fiber-coupled device. Additionally, we use a commercially available and quickly reactive epoxide for curing, which is developed for fiber connector preparation and is thus index matched to optical fibers. The index matching between fiber-core and adhesive further reduces reflection losses at the interface which can further enhance the extraction efficiency. Moreover, our approach can easily be adopted for other solid-state based SPSs, such as micropillar cavities \cite{Heindel2010} and bullseye resonators \cite{Ates2012a}, as well as for SPSs emitting at telecom wavelengths. 

\subsection*{Stand-alone single-photon unit}

\begin{figure}[t]
\centering
\includegraphics[width=\linewidth]{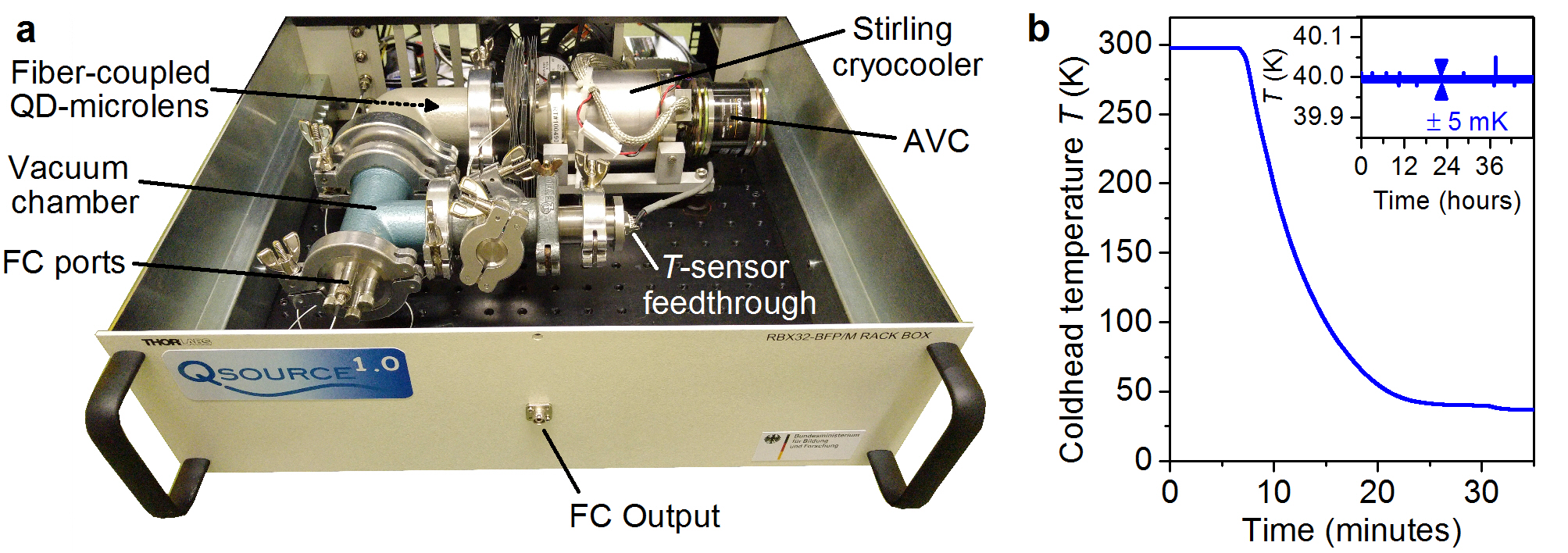}
\caption{The stand-alone single-photon source 'QSource'. (a) QSource module comprising the Stirling cryocooler with attached customized vacuum chamber. The fiber-coupled (FC) QD-microlens (cf. Figure \ref{Fig_1}) is mounted to the cryocooler's coldhead inside the vacuum chamber. The active vibration cancellation (AVC) system reduces vibration export by the cryocooler's moving piston. (c) Coldhead temperature of the Stirling cryocooler measured during cool-down. Inset: Coldhead temperature over a measurement period of 48\,hours, revealing high temperature stability of the cryocooler.}
\label{Fig_2}
\end{figure}

When operated at cryogenic temperatures, the InAs/GaAs material system used for the QD-SPS in our work offers superior quantum optical properties \cite{Michler2000,Ding2016} if compared to quantum emitters operated at room temperature \cite{Michler2000a,Kurtsiefer2000,Lounis2000,Holmes2014,Deshpande2014}. Thus, a crucial aspect for the development of high-quality and user-friendly SPSs is the cooling of the quantum emitter to cryogenic temperatures. To date, liquid helium supplied in large storage dewars in combination with flow-cryostats is still the most common cooling technique in research laboratories. However, aiming at the autonomous operation of a quantum light source at remote places without laboratory infrastructure requires more practical cryocoolers. In recent years the use of closed-cycle pulse-tube coolers for experiments in quantum optics have become more popular, due to improvements in vibration damping \cite{Rau2014}. Although such systems do not require a permanent liquid helium supply, they still depend on high-power voltage supplies and bulky compressors. 

In our work, we employ a compact and economic Stirling cryocooler \cite{Veprik2005} to build a stand-alone fiber-coupled SPS (see Figure \ref{Fig_2}\,a), called 'QSource' in the following. The Stirling cryocooler only requires a standard supply voltage (220\,V), can be operated with minimal space requirements and recently proved to be suitable for quantum-optical free-space experiments at base temperatures down to 30\,K \cite{Schlehahn2015}. The Stirling cryocooler (Model: Cryotel-GT from Sunpower) used in our QSource comprises a single piston moving along the same axis as the movable regenerator which allows for compact dimensions of $27.6\,\rm{cm}\times8.3\,\rm{cm}\times8.3\,\rm{cm}$. The cryocooler is additionally equipped with an active vibration cancellation (AVC) system. The AVC constitutes a dynamically (real-time) adjusted inert mass, which minimizes the vibrations exported by the cryocooler's moving piston. For this purpose, the vibrations are measured using an accelerometer based on micro-electro-mechanical systems (MEMS) mounted at the housing of the Stirling cryocooler. The resulting signal is used for an active feed-forward control loop for adjusting the balancers movement (see Ref. \cite{Riabzev2009} for details). The fiber-coupled SPS described in the previous section is mounted directly to the cryocooler's coldhead. A custom made vacuum chamber attached to the Stirling cryocooler encloses the coldhead and contains ports for up to four optical fibers and the electrical-feedthroughs for the temperature sensor integrated into the coldhead.
 
Due to the compact dimensions of the Stirling cryocooler, the complete single-photon unit including the vacuum chamber with all necessary feedthroughs fits easily within a standard 19$"$ rack-insert as depicted in Figure \ref{Fig_2}\,b). For an autonoumous operation of our QSource, the single-photon unit is housed in a small mobile rack (90\,cm height) containing a diode laser ($\lambda=651\,$nm or 855\,nm, continuous wave (CW) or pulsed mode) coupled to the single-photon unit via a 90:10 fiber beamsplitter, the electronics for controlling the Stirling cryocooler and a personal computer (PC) with display, keyboard and mouse. Via the PC the user is able to control all parameters relevant for operation of the QSource, including temperature control and monitoring. Additionally, it optionally allows for photon-autocorrelation measurements using a fiber-based Hanbury-Brown and Twiss (HBT) setup equipped with two Silicon-based single-photon counting modules (SPCMs) and the corresponding electronics for time-correlated single-photon counting (TCSPC). The temperature at the coldhead during a cool-down of the Striling-cryocooler is depicted in Figure \ref{Fig_2}\,c. After only 30\,minutes the sample reaches a temperature of 40\,K. The inset shows a measurement of the coldhead temperature over a period of 48\,hours, revealing a  mean base temperature of 39.995\,K at a temperature stability of $\pm6\,$mK being competitive with state-of-the-art Helium-flow cryostats and about one order of magnitude improvement compared to our previous work Ref. \cite{Schlehahn2015}. The achieved temperature is sufficiently low for applications relying solely on the sub-Poissonian statistics of the single-photon states generated by QD emitters, such as quantum key distribution via the BB84 protocol \cite{Heindel2012}. Quantum information schemes relying on the photon indistinguishability, however, require temperatures below 10\,K \cite{Thoma2016}, which can be reached in compact devices in future by combining Stirling- and Joule-Thomson- cryocoolers \cite{Gemmell2017}.

\subsection*{Durability of fiber-coupling}

\begin{figure}[t]
\centering
\includegraphics[width=\linewidth]{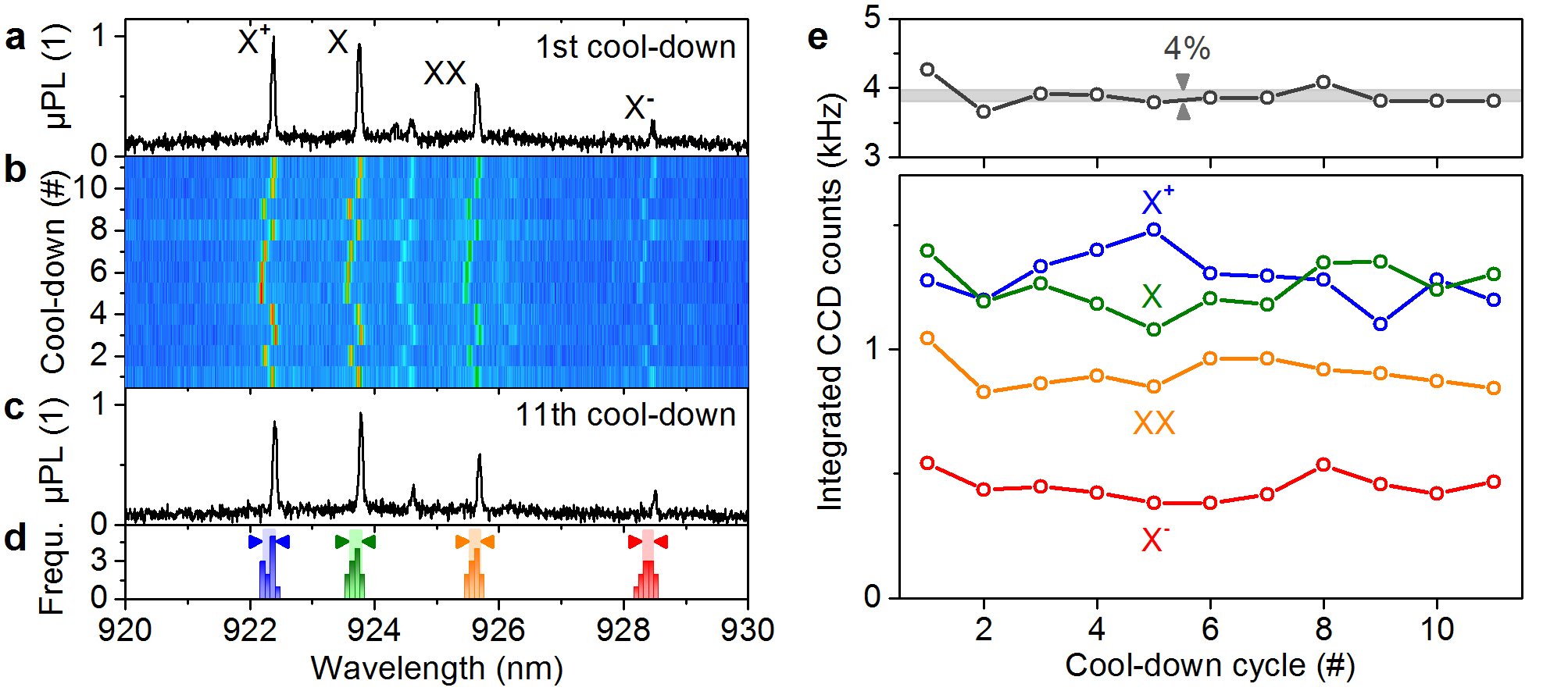}
\caption{Durability test of the fiber-connection. (a), (b) and (c) Spectra of a fiber-coupled QD-microlens (Device1) operated in our QSource at $T=40\,$K after repeated cool-down/warm-up (40\,K\,$\leftrightarrow$\,290\,K). Emission of the exciton (X), the biexciton (XX), and singly charged trion states (X$^+$ and X$^-$) is identified. (d) Frequency count histogram of the center wavelength of the four QD states extracted from the spectra in the contour plot in (b). (e) Integrated CCD counts of the fiber-coupled QD emission for each individual excitonic complex (lower panel) and its sum (upper panel), evaluated for the spectra shown in (b).}
\label{Fig_3}
\end{figure}

The spectral properties and the durability of our fiber-coupled SPS are tested using a spectrometer with a 1200-lines optical grating and an attached liquid-Nitrogen-cooled charge-coupled device camera (system's spectral resolution: 25\,$\mu$eV). Figure \ref{Fig_3}\,a presents the spectrum of a fiber-coupled QD-microlens, named Device1 from now on, operated in our QSource at 40\,K (integration time: 1\,s). Emission of different excitonic states stemming from the same QD is clearly observed, where the charge-neutral exciton (X), the biexciton (XX) and the singly charged trion states (X$^+$ and X$^-$) were identified via polarization- and excitation-power-resolved measurements (not shown). 

To evaluate the durability of the glued connection between optical fiber and QD sample, we repeatedly warmed up and cooled down the sample and monitored the emission of Device1 using an automated routine as described in the following. Starting from room temperature (290\,K), the Stirling cryocooler is turned on in power-controlled mode (110\,W). After reaching the base temperature, the system holds this temperature for 30\,minutes to reach thermal equilibrium. Then the coldhead temperature is stabilized at 40\,K using a proportional-integral-derivative (PID) -controlled operation mode for 10\,minutes and a spectrum is recorded before the Stirling cryocooler is turned off. The next cooling cycle starts automatically after the coldhead reached room temperature again (about 6\,hours). Figure \ref{Fig_3}\,b displays the spectra of Device1 in a contour plot for cool-down/warm-up cycles 1 to 11, where the last spectrum is additionally shown in Figure \ref{Fig_3}\,c. The emission properties of the QD in terms of the spectral finger print and the emission intensities of the corresponding excitonic states remain almost unchanged, indicating a high durability of the glued connection between fiber and sample. Quantitatively extracting the emission energies of the four QD states for each spectrum from cool-down cycle 1 to 11 by fitting yields standard deviations for the spectral shifts of $\delta E_{\rm{X+}}=\pm86\,$pm, $\delta E_{\rm{X}}=\pm85\,$pm $\delta E_{\rm{XX}}=\pm84\,$pm and $\delta E_{\rm{X-}}=\pm98\,$pm. The corresponding frequency histogram of the fitting results is shown in Figure \ref{Fig_3}\,d. The small deviation in emission energy over many cool-down cycles is attributed to slightly varying electric fields in the vicinity of the QD due to charge fluctuations and the resulting quantum confined Stark effect \cite{Empedocles1997}. Interestingly, a careful analysis of the data reveals that the relative spectral positions of all four emission lines with respect to the X-state remains constant within a standard deviation of $\pm4$ to $\pm7\,$pm, more than one order magnitude less than the absolute spectral shift. Thus we conclude, that the binding energies of the respective QD-states remain almost constant over the entire eleven cool-down cycles, implying that the strain inside the QD sample is not changing over time, which confirms the high stability of our fiber-coupling scheme. Additionally, we evaluate the change in brightness of Device1 during the durability test. Figure \ref{Fig_3}\,e depicts the integrated CCD counts of the QD emission for each individual excitonic complex (lower panel) and its sum (upper panel), evaluated for the spectra shown in Fig. \ref{Fig_3}\,b. We observe, stable emission of Device1 with a standard deviation of $4\%$, certifying an excellent durability of the glued fiber connection used in our QSource. The noticeable anti-correlation between the emission intensities of X and X$^+$ state in the lower panel indicate a random positive charging of the QD by nearby traps.

Beyond the thermal durability proven above, real-world applications of our QSource will also demand robustness against mechanical shocks during e.g. the shipment or the operation on moving platforms such as airplanes. In this context, it is worth mentioning that the repeated transfer between remote labs on different floors in our institute had no notable influence on the performance of the QSource, which was integrated in a wheeled measurement rack for this purpose. This underlines the high mechanical durability of the bond between fiber and QD-sample, which can be analyzed in more detail by performing quantifiable shock tests in future.

\subsection*{Turn-key single-photon generation}

\begin{figure}[t]
\centering
\includegraphics[width=\linewidth]{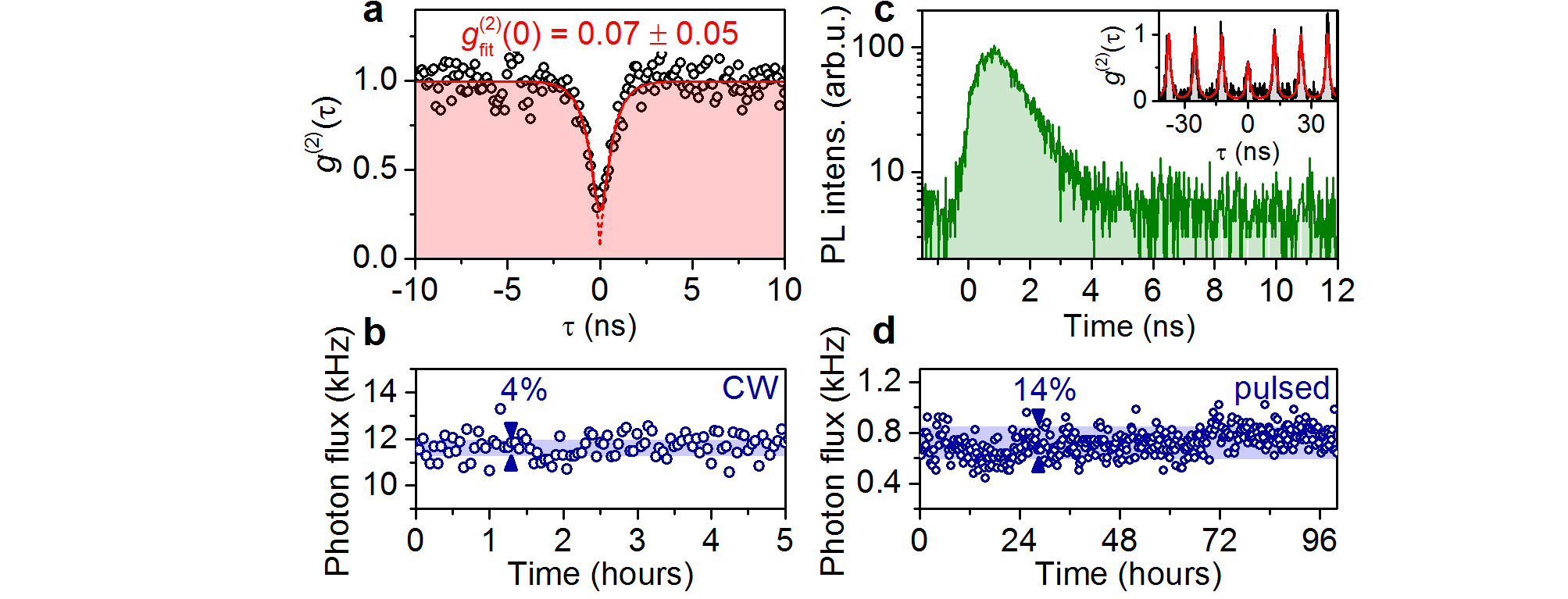}
\caption{Single-photon generation using the QSource. (a) Measurement of the second-order photon-autocorrelation $g^{(2)}(\tau)$ on the X$^+$-emission of Device2 (cf. Figure \ref{Fig_3}) operated in the Stirling cryocooler demonstrating single-photon emission with $g^{(2)}(0)=0.07 \pm 0.05$. The solid (dashed) line represents the convoluted (deconvoluted) $g^{(2)}(\tau)$-function of a model fit taking into account the setup's temporal resolution. (b) Photon flux of Device2 recorded at the single-photon counting modules during the $g^{(2)}(\tau)$-measurement shown in (a). (c) Time-resolved measurement of the fiber-coupled emission of Device3 under pulsed excitation at 80\,MHz. Inset: $g^{(2)}(\tau)$-histogram recorded for Device3, revealing triggered non-classical light emission. (d) Photon flux of Device3 under pulsed excitation during a user-intervention-free test run over a period of 100\,hours.}
\label{Fig_4}
\end{figure}

To evaluate the performance of our stand-alone single-photon source in terms of its single-photon purity, we performed measurements of the second-order photon-autocorrelation $g^{(2)}(\tau)$ during operation of the Stirling cryocooler at cryogenic temperature ($T=40\,$K). For this purpose, we evaluated another fiber-coupled single-photon source (Device2), which featured particular bright emission of the positively charged trion state X$^+$. The fiber-output of the QSource is spectrally filtered (bandwidth: 0.1\,nm) via the external spectrometer transmitting the emission of the X$^+$ state. Right after the spectrometer, the emission is coupled to the fiber-based HBT setup for coincidence measurements. Figure \ref{Fig_4}\,a presents the histogram of $g^{(2)}(\tau)$ (time-bin width: 275\,ps) under CW excitation at 651\,nm. The pronounced antibunching at zero time delay ($\tau=0$) signifies the non-classicality of the emitted light. Indeed, fitting the experimental data by accounting for the timing resolution (350\,ps) of the HBT setup reveals $g^{(2)}(0)=0.07 \pm 0.05$ unambiguously demonstrating single-photon emission. Additionally, we evaluated the photon flux of Device2 during the $g^{(2)}(\tau)$-measurement as shown in Figure \ref{Fig_4}\,b. We observe a mean single-photon flux at the HBT setup of 11.7\,kHz at a standard deviation of only 4\% during the measurement.

With respect to applications of single-photon sources in quantum information, triggered emission of single-photons is required. We therefore tested the QSource under pulsed-excitation at 855\,nm with an excitation repetition rate of 80\,MHz (pulse-width: 80\,ps) using yet another fiber-coupled microlens (Device3) acting as quantum emitter. The choice of Device3 was necessary, because Device2 showed a long decay-time constant ($>10\,$ns) under pulsed-excitation, most probably due to pronounced charge-carrier recapture processes \cite{Aichele2004}. The time-resolved fiber-coupled emission of an excitonic state is displayed in Figure \ref{Fig_4}\,c. The optical response of the QD shows a monoexponential decay with a lifetime of $(1.0\pm0.1)\,$ns. The corresponding $g^{(2)}(\tau)$-histogram under pulsed excitation is depicted in the inset of Figure \ref{Fig_4}\,c. The clearly separated coincidence peaks combined with the significantly suppressed peak at zero delay prove triggered emission of non-classical light. Fitting the experimental data with a model according to Ref. \cite{Schlehahn2016} yields antibunching with $g^{(2)}(0)=0.57\pm0.05$. Here, the non-ideal antibunching is limited by the uncorrelated background emission of the nearby wetting layer for this particular device. Additionally, we also evaluated the long-term stability of our QSource for Device3 (see Figure \ref{Fig_4}\,d). We observe a photon flux at the HBT setup of 720\,Hz (including 40\,Hz dark counts) with a standard devication of $14\%$ during the entire 100-hour measurement period without any user intervention, again confirming the excellent stability of our QSource.

To assess the photon collection efficiency of our devices, we determined the overall transmission of our complete experimental setup to be about 0.3\% (from the fiber inside the QSource to the SPCMs). Hence, the 680\,Hz of detected QD-photons in the pulsed experiment correspond to a photon flux of 227\,kHz inside the fiber and a collection efficiency of 0.28\% (=227\,kHz/80\,MHz) per excitation pulse. Similarly, in case of the CW measurement, 11.7\,kHz correspond to 3.9\,MHz photon flux inside the fiber and a collection efficiency of 0.39\% (=3.9\,MHz/1\,GHz) per excitation (considering a radiative lifetime of 1\,ns). For Device2, we independently measured a photon extraction efficiency of 14.4\% into $NA= 0.4$ (in a free-space experiment), which is consistent with device simulations based on finite element calculations (see Supplementary of Ref. \cite{Gschrey2015b}). Hence, we estimate a fiber-coupling efficiency of 2.7\% for Device2 used for the measurements in Fig. \ref{Fig_4}\,a and b. We anticipate significantly improved coupling efficiencies into optical fibers, by employing recently developed on-chip micro-optics \cite{Fischbach2017a} in future QSource modules.

\section*{Summary and outlook}

In this work, we report on a user-friendly stand-alone SPS providing single-photon emission via an optical fiber. Within our single-photon unit, a QD is deterministically embedded in a monolithic microlens. The precise coupling of the QD-microlens to the optical fiber is achieved using a robust process based on active alignment and epoxide adhesive bonding at room-temperature, which allows for an accuracy better than 9\,$\mu$m. The fiber-coupled single-photon emitter is mounted to the coldhead of a compact plug-and-play Stirling cryocooler integrated in a 19$"$ rack-insert. Our stand-alone device allows for autonomous operation over several days with high stability of the single-photon flux at the fiber output and antibunching values down to $g^{(2)}(0)=0.07 \pm 0.05$. Moreover, we show triggered emission of non-classical light at 80\,MHz excitation repetition rate. The high durability of the fiber-connection is proven in endurance tests, revealing stable integrated QD emission within 4\% over eleven successive cool-down/warm-up cycles. Additionally, a user-intervention-free 100-hour test run proves the long-term stability of our QSource and we achieve single-photon detection rates of up to 11.7\,kHz at a standard deviation of 4\%, confirming the potential of our approach for applications in quantum information science.

By implementing straight forward extensions to our QSource, we anticipate greatly enhanced performance in terms of the achievable single-photon purity and flux. For instance, by combining our deterministic QD-microlenses with miniaturized laser-written multi-lens objectives \cite{Gissibl2016}, the photon extraction efficiency from our device as well as the coupling efficiency to an optical fiber can be enhanced \cite{Fischbach2017a}. 
Moreover, a higher level of integration will reduce the total system losses. This can for instance be achieved by using electrically pumped quantum light sources by utilizing a recently developed contacting scheme for deterministic QD microlenses \cite{Schlehahn2016a} or by employing compact interference band-pass filters instead of bulky external spectrometers \cite{Heindel2012}. Additionally, the accuracy of our fiber-coupling process also shows promise for the use of single-mode fibers, which however will require further optimization. Finally, the next-generation QSource can easily be adapted to other operation wavelengths, opening up the route for the realization of turn-key stand-alone quantum light sources at telecommunication wavelengths for applications in fiber-based quantum key distribution and quantum repeater networks.

\section*{Acknowledgements}

We thank Jan Hausen for technical assistance and acknowledge financial support from the German Federal Ministry of Education and Research (BMBF) via the VIP-project QSOURCE (Grant No. 03V0630), the German Research Foundation (DFG) via the SFB 787 'Semiconductor Nanophotonics: Materials, Models, Devices', the European Research Council under the European Union's Seventh Framework ERC Grant Agreement No. 615613, and the project EMPIR 14IND05 MIQC$^2$ (the EMPIR initiative is co-funded by the European Unions Horizon 2020 research and innovation programme and the EMPIR Participating States).

\section*{Author contributions statement}

A. Schlehahn and T.H. developed the stand-alone single-photon source, performed the spectroscopy and correlation experiments together with S.F. and analyzed the experimental data. S.F. performed the CL lithography and processed the samples together with R.S. under supervision of S. Rodt. A.K. and A. Strittmatter grew the samples. T.H. wrote the manuscript with input from all authors. S. Reitzenstein supervised the project.



\end{document}